\documentclass[twocolumn,showpacs,preprintnumbers,amsmath,amssymb]{revtex4}
\usepackage{graphicx,epsf}
\begin{document}

\title{An analytic model for a cooperative ballistic deposition in one dimension}
\author{M. Kamrul Hassan$^{\dagger,\ddagger}$, Niels Wessel$^\dagger$, and J\"urgen Kurths$^\dagger$}
\affiliation{
$^\dagger$University of Potsdam, Department of Physics, Postfach 601553, D-14415 Potsdam, Germany \\
$^\ddagger$ University of Dhaka, Department of Physics, Theoretical Physics Division, Dhaka 1000, Bangladesh
}
%\email{khassan@agnld.uni-potsdam.de}
\date{\today}

\begin{abstract}%

We formulate a model for a cooperative ballistic deposition (CBD) process whereby the incoming particles are 
correlated with the ones already adsorbed via attractive force. The strength of the correlation is controlled by a 
tunable parameter $a$ that interpolates the classical car parking problem at $a=0$, the ballistic deposition at $a=1$ and
the CBD model at $a>1$. The effects of the correlation in the CBD model are as follows. 
The jamming coverage $q(a)$ increases with the strength of attraction $a$ due to an ever increasing tendency of cluster 
formation. The system almost reaches the closest packing structure as $a\rightarrow\infty$ but never forms a 
percolating cluster which is typical to 1D system.  
In the large $a$ regime, the mean cluster size $k$ increases as $a^{1/2}$. Furthermore, 
the asymptotic approach towards the closest packing is purely algebraic both with $a$ as $q(\infty)-q(a) \sim a^{-1/2}$
and with $k$ as $q(\infty)-q(k) \sim k^{-1}$ where $q(\infty)\simeq 1$.
\end{abstract}


 \maketitle

\section{Introduction}

The kinetics of a monolayer growth by the deposition of macromolecules and colloidal particles onto solid substrates 
has been the subject of extensive research for the past years (see \cite{kn.evans,kn.tarjus,kn.schaaf} for extensive
review). The reason is well justified because its importance 
and significance cover many seemingly unrelated topic in physics, chemistry, biology and other branches of 
science and technology. From a theoretical point of view, the random sequential adsorption (RSA) of a 
monodisperse particle is one of the simplest model that can describe deposition phenomena \cite{kn.renyi}. 
In this process, particles are deposited randomly, one at each time step, with the strict restriction that overlapping is forbidden. This can be described by the 
following algorithm. (i) At each time step, a random position is chosen from the whole substrate and is assigned to the 
center of the particle picked for deposition. (ii) If the incoming particle collides with a previously adsorbed one, 
the trial attempt
is rejected; otherwise it is adsorbed irreversibly. (iii) In either case, the time is increased by one unit and the steps
(i) - (ii) are repeated until the system reaches a state when particles can no longer be adsorbed.

One of the virtue of the RSA model is that like many statistical physics problems, it is exactly solvable in one dimension 
in both its continuum and lattice version for some specific cases. The one dimensional continuum version of the model is 
popularly known as the {\it random car parking} (RCP) problem and has attracted much attention.
Despite the inherent simplicity in the RSA model, it still captures essential generic features of the process and 
has proved to describe successfully the behavior of many experimental systems, namely, the adsorption of proteins, 
latex and colloidal particles \cite{kn.feder,kn.onoda,kn.mann}. Nevertheless, there have been continuous research 
efforts to include various important physical features to make it more realistic and thus covering a wider range of 
real life situations \cite{kn.micha,kn.pagonabarraga,kn.rodgers}.
Along this road, a good deal of progress has already been achieved and yet we are far away from a complete theory.
In recent years it has received extra momentum and the number of papers published in the recent years is a clear 
testimony to it \cite{kn.schaaf,kn.hassan,kn.bonnier,kn.pastor,kn.lee,kn.penrose}.

The strongest criticism of the RSA model is its outright rejection of particles that fall upon an already adsorbed one. 
The most successful model overcoming this criticism is known as the ballistic deposition (BD) 
process \cite{kn.talbot,kn.jullien}. The BD model is best explained in terms of the deposition of disks of 
diameter $m$ on a line. In this case, whenever an incoming disk overlaps with an already adsorbed one, it is allowed 
to roll over the latter disk following the path of the steepest descent. In doing so, the disk can either touch the 
adsorbing plane (global minimum) or it may 
find itself trapped in the local minimum formed by two or more connected disks.
In the former case, the disk is irreversibly attached with the one it rolled over leaving no gaps
in between, while in the latter case the trial attempt is rejected.
Both in the simple RSA and the BD model, only a short range hard-core repulsion via the excluded volume 
effects is taken into account. All forms of  
long-range interactions between the particles in the adsorbed and adsorbing phases are ignored.
There are some fragmented attempts though, 
to include some specific forms of interactions such as the electrostatic, dipolar and the hydrodynamic interactions
\cite{kn.electrostatic,kn.dipolar,kn.hydrodynamic}.

In this article, we consider a model that includes the 
attractive force between the particles in the adsorbing and the adsorbed phase mimicking the long range interaction. 
To study the underlying mechanisms 
in such  complex phenomena, like the deposition processes, it is of great advantage to have 
a flexible model which allows explicit analytical solutions with different possible mechanisms.
In order to increase the flexibility of the model,
we further generalise it by introducing a parameter $a$ that can tune the strength of 
the attractive forces. This would certainly facilitate the study of the general effect of long-range interactions in
the whole process. We can recover the simple RSA results by setting $a=0$, the BD results by setting $a=1$ and while  
$a>1$, we show that the model describes the cooperative ballistic deposition (CBD). Obviously, the strength of correlation
is determined by the strength of the attractive force. This results in an increased overlapping with the already adsorbed 
particles. However, once a particle collides with an already adsorbed one, it follows the rule of the simple BD. 
Thus, as the strength of interaction increases, we expect an increasing rate of successful adsorptions.
The analytical part of the model we study in this work is similar to the one previously studied by Viot {\it et al} 
\cite{kn.viot}. However, we go far beyond that not only by offering a completely new model but also by giving
an in-depth analysis and an exact physical explanation as well as extracting more insightful results from
several aspects. First, we give
a general explanation of the model and a means to convert it into a simple and well known BD model based on a philosophical
approach. Our ideas are backed up and well supported by direct numerical simulations which 
enables to understand the physical nature of the system described by each term of the rate equations we used for 
analytical solutions. Second, we present more extensive results showing the 
asymptotic approach of the coverage towards the jamming limit in terms of the various parameters involved in the processes.

%\begin{eqnarray}  
%\label{eq:gap_1}
%{{\partial c(x,t)}\over{\partial t}} & = &- (x-\bar\sigma)c(x,t) +  q\a \sigma \Big[c(x+\sigma,t)-c(x,t)\Big] 
%\nonumber \\ && + p\beta m\sigma\Big [c(x+m\sigma,t)-c(x,t)\Big] \nonumber \\ &&
% 2q\int_{x+\sigma}^\infty c(y,t) dy + 2p\int_{x+m\sigma}^\infty c(y,t)dy
%\end{eqnarray}

\section{cbd model}

We consider a system which consists of a reservoir of particles with diameter $m$ lying in the immediate vicinity of 
a substrate which is an infinitely extended line. 
The adsorbing particles may be in the gas or in the fluid phase and arrive in the adsorbing plane through
Brownian motion. As soon as a particle comes into contact with a gap large enough to accommodate it, it 
is then adsorbed immediately and irreversibly. In addition, the incoming particle that touches an already adsorbed 
one, is allowed to follow the BD rules to form a monolayer. The simplicity of the 1D problem lies in the following 
situations. Every successful deposition of a particle on a given gap divides it into smaller gaps having the same 
geometry as the parent gap. It is this {\it shielding property}, found only in 
1D, which we shall use to gain further insight into the problem and tackle it analytically. For simplicity's sake, we assume 
that the daughter gaps are uncorrelated, irrespective of the island size separating the gaps from their neighbours, 
so that we can treat each gap as an independent entity. We further assume that each roll-over motion is
completed prior to the next trial attempt for deposition. 

\begin{figure}[!htb]
\centerline{\includegraphics[width=8.0cm]%
{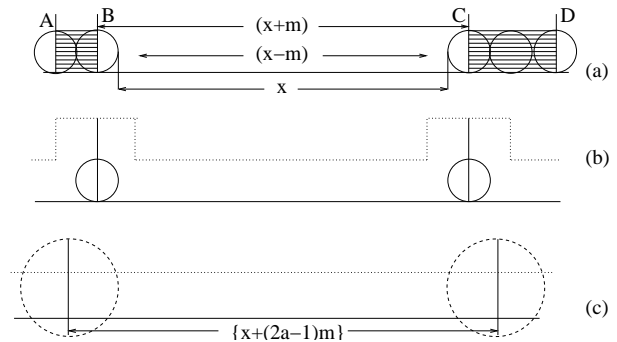}}
 \caption{Schematic illustration of the model in different situations. 
The shaded regions between $AB$ and $CD$ in (a) have local minimums and play no role except for kinetic reasons. 
We therefore eliminate all such shaded regions so that we have a system where all the gaps are separated 
from their neighbours by only one disk as shown in (b). The dotted lines in (b) represent the probability 
distribution in different regimes of a given gap.
The steepest descent path in (c) is artificially increased by using the
idea of a virtual disk at the expense of lowering the height of PD so that every point of $x+(2a-1)m$ is now
equally likely to be chosen by the trial attempt.}
\label{fig:1}
\end{figure}

At this point, it is useful to discuss the classical RSA and the simple BD model before introducing the
cooperative BD model. In the classical RSA, an incoming particle is 
adsorbed successfully in a gap of size $x$, if the center of the incoming particle arrives in any place but $m/2$ 
away from either edges of the gap. This means that only $(x-m)$ of a given gap $x$ is accessible for adsorption, 
which we have illustrated in Fig. 1(a). In the BD model, on the other hand, those particles that fall on an already 
adsorbed one may reach the substrate via the roll over motion. The deposition via rolling is successful if 
the center of the incoming particle falls within a distance of $m/2$ on either sides of both edges. 
It is then adsorbed on the respective edge creating a 
new gap of size $(x-m)$. That is, for a given gap $x$, the total position accessible to a new arrival is $(x+m)$ which is
shown in Fig. 1(a). Note that any particle dropping in the shaded regimes $AB$ or $CD$ 
are considered to be trapped due to the local minimum and will never reach the global minimum or the adsorbing substrate. 
Thus, if we are not interested in the kinetic aspect of the process, we can safely delete the shaded regimes as if they
did not exist and assume that the neighbouring gaps are separated by only one disk as shown in Fig. 1(b).    
We can thus define each gap as an independent isolated interval bordered on either end by a semidisk so that if 
we had connected the two remote ends it would then form a ring with one particle at the joint.

We are now in a position to introduce the long range attractive force among the incoming particles and the 
particles in the 
adsorbed phase. The question is how to incorporate it? First, we need to understand the effects of such an
attractive force. The most significant one is that each adsorbed particle 
will tend to attract the incoming particle towards it. This immediately breaks the random nature of the process as the 
incoming particles are more likely to land on an already adsorbed particle than on the gap. That is, as far as the
selection of the trial position is concerned, we have two distinct probability distribution (PD)
regimes as indicated in Fig 1(b). First, the {\it force free regime} where the attractive force is strictly zero. 
Second, the {\it force field regime} where the 
adsorbed particle exerts force and hence can influence the selection of the trial position. However, for analytical 
simplicity, we assume a square well potential around the center of each adsorbed particle of width $2m$. That is,
the strength of the attractive force is equal anywhere within the force field regime. This is 
indicated by the dotted lines in Fig. 1(b) representing the PD of the position chosen by the trial attempt. 
The flat PD implies that all the points within that regime have the same {\it a priori probability}.

\begin{figure}[!htb]
\centerline{\includegraphics[width=8.0cm]%
{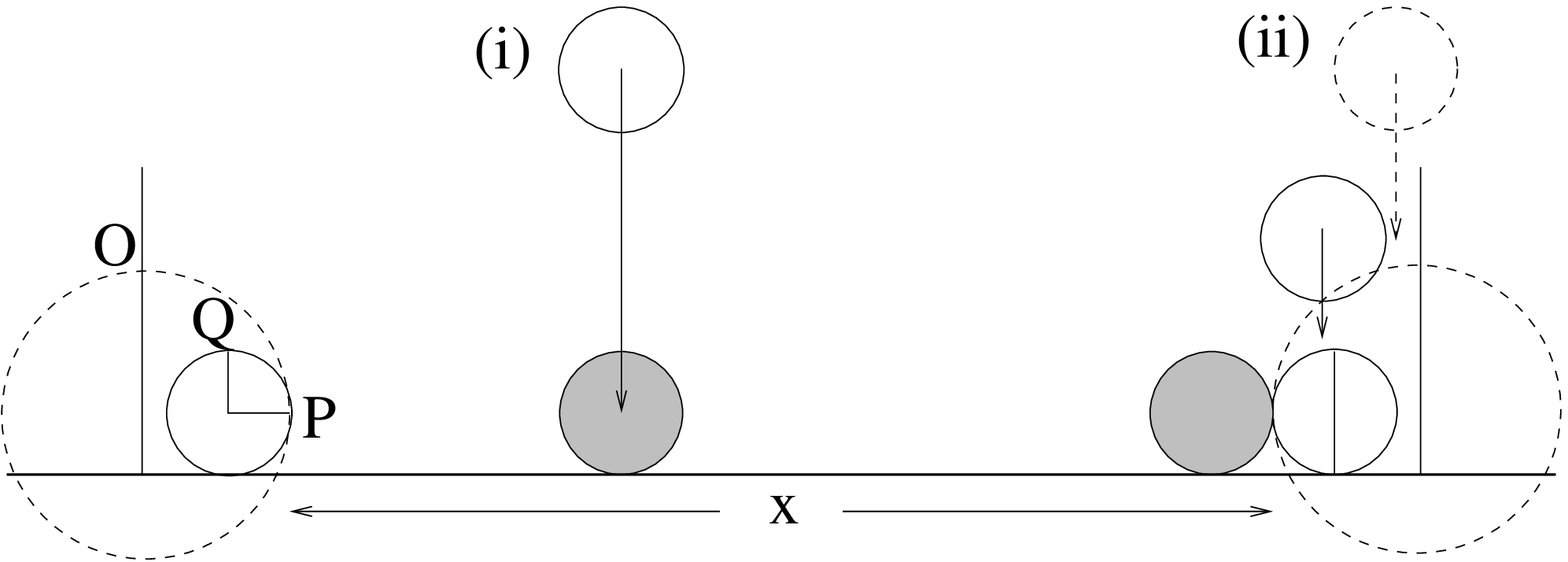}}
 \caption{Schematic illustration of the rules for the cooperative ballistic deposition model. 
The incoming disk is adsorbed directly as shown in (i) if it is dropped  within $(x-m)$. The deposition via the rolling 
mechanism is depicted in (ii). Here, any disk that falls on the steepest descent path of the virtual disk, OP, 
is assumed to be dropping on its corresponding equivalent point of the real particle QP from where it can 
successfully reach the global minimum.}
 \label{fig:2}
\end{figure}

It is noteworthy to mention that the simple BD model 
refers to the case where  both regimes have the same flat PD height and hence the whole substrate represents
the zero force field or zero attractive force. Now, as soon as we switch on the attractive force, the height of the 
PD around each adsorbed particle will increase to a degree depending on the strength of the attractive force. However, we can lower the PD height
by increasing the width keeping the total area unchanged. In this way, we can make the whole system having the 
same PD height and treat it like the simple BD model where an incoming particle can land anywhere in the substrate
with equal probability including the disks that are already occupied as represented in Fig. 1(c). 
That is, we can envision it as follows: 
Prior to selecting a position for an incoming particle, we replace the size of the depositing particles used for
defining the gaps by a virtual disk of diameter $R>m$, without altering the gap size. In this way, we increase the 
cross-section of collision for an incoming particle in order to count the additional collision probability due to
the attractive force. Once the position for the next trial deposition is chosen, 
we can immediately return to the system with disks having diameter $m$ and  proceed according to the simple BD rules
as depicted in Fig. 2.
In doing so, we artificially increase the probability with which an incoming particle may collide with an already 
adsorbed disk thus enhancing the probability of adsorption via the rolling mechanism and mimicking the effect of 
the attractive force. One can thus expect an enhanced adsorption probability near the 
two extreme ends of each gap as the virtual disk size increases and, in the limit $R\longrightarrow \infty$, we can only
expect the adsorption via the rolling mechanism except in the very early stage where the virtual diameter and the gap size
may be of the same order in size. 
%The present model has the further very intuitive interpretation.
% We can envision such process as the effect of long-range attractive interactions 
%that eventually tend to attract the incoming particles towards the  already adsorbed  particles. 
%The strength of attraction in this model is measured in terms of the virtual disk size $R$. 

\section{Analytical solution}

To address the problem described above analytically, 
we adopt the well studied rate equation approach of the gap size distribution function or concentration 
$c(x,t)$. Let us define $c(x,t)dx$ as the 
number of gaps at time $t$ in the size range between $x$ and $x+dx$. 
The kinetics of adsorption of the monodisperse particles can then be described by the following set of rate equations:
\begin{eqnarray}
{{\partial c(x,t)}\over{\partial t}}& = &-(x-m)c(x,t)+2 \int_{x+m}^\infty c(y,t)dy \nonumber \\ && +
2am \Big \{c(x+m,t)-c(x,t)\Big\}
\end{eqnarray}
for $x\geq m$ and
\begin{equation}
{{\partial c(x,t)}\over{\partial t}} = 2 \int_{x+m}^\infty c(y,t)dy +2am c(x+m,t)
\end{equation}
for $x<m$. The above rate equations are mean-field in nature as the fluctuations and correlations are ignored. 
The rate equation approach is 
based on the assumption that the creation and annihilation of gaps are independent of the size 
of the neighboring gaps. The first two terms of Eq. (1) and the first term of Eq. (2) are the same as that of the simple 
RSA process and thus describe the creation and destruction of a gap of size $x$ due to the direct 
adsorption of size $m$ on size $y\geq x+m$ or on size $x$, respectively. The remaining terms in both equations also 
describe the creation and destruction of gaps but due to the rolling motion over the steepest descent path 
to travel the maximum linear path $am$, where $a$ is a dimensionless constant number that we can tune. 
The factor `2' in the integral terms accounts for
the fact that any of the two new gaps created upon a direct deposition on the gap size  $y\geq x+m$ can be of
size $x$; whereas the same factor in the remaining terms takes into account that a gap of size $x$ can be 
created or destroyed from either end by adsorption. In order to understand the role of $a$, 
it is convenient to rewrite Eq. (1) as
\begin{eqnarray}
{{\partial c(x,t)}\over{\partial t}}& = &-\{x+(2a-1)m\}c(x,t)+2 \int_{x+m}^\infty c(y,t)dy \nonumber \\ && +
2amc(x+m,t).
\end{eqnarray}

The term $\{x+(2a-1)m\}$ in the above equation is the key to understand the role of $a$. Note that by setting
$a=0$ we recover the classical RSA case where $(x-m)$ of a given gap $x$ is accessible for adsorption, which is
consistent with our discussion in the previous section. Similarly, $a=1$ describes the simple BD model where the 
total positions accessible to a new arrival is $(x+m)$ which is again consistent (see Fig. 1). 
In general, $\{x+(2a-1)m\}$ means that a given gap of size $x$ is bounded by a semidisk
of diameter $R=(2a-1)m$ while the adsorbing particles are of size $m$. That is, any particle that falls within a distance
$am$ from either end of  $\{x+(2a-1)m\}$ effectively will collide with the virtual disk. Every point of the steepest 
descent of the virtual path $OP$ in Fig. 2 has its corresponding equivalent point on the real path $QP$. Therefore, an 
incoming particle falling on the virtual path is assumed as if it were falling on the exact equivalent position 
of the real path and vice versa.

To solve Eq. (1) we seek a trial solution of the following form  
\begin{equation}
c(x,t)=A(t)e^{-(x-m)t},
\end{equation}
 where $A(t)$ is still an undetermined quantity fixed by the initial condition. Let us assume a monodisperse initial
condition $c(x,0)=\delta(x-L)/L$ so that we have
\begin{equation}
 \lim_{L\longrightarrow \infty}\int_0^L c(x,0)dx =0, \hspace{0.3cm} \lim_{t \longrightarrow 0}\int_0^\infty xc(x,t)dx=1.
\end{equation}
Substituting the trial solution into Eq. (1), 
we obtain the following differential equation for $A(t)$
\begin{equation}
{{d\ln A(t)}\over{dt}}={{2e^{-m t}}\over{t}}+2ae^{-m t}.  
\end{equation}
Solving it, satisfying the initial conditions, we get
\begin{equation}
A(t)=t^2F(a, m t)
\end{equation}
where the auxiliary function $F(a,m t)$ is defined as
\begin{equation}
F(a, m t)= e^{-2\int_0^{m t}{{1-e^{-u}}\over{u}}du+2a(1-m t -e^{-m t})}.
\end{equation}
To obtain $c(x,t)$ for $x<m$, we substitute the solution of Eq. (1) into Eq. (2) and then upon a direct integration 
we get
\begin{equation}
c(x,t)=\int_0^t u(2+ am u)F(a,m u)e^{-xu}du.
\end{equation}
The solutions $c(x,t)$ can provide a complete analytical description of the process including its kinetic aspect.
All we need now is to find
useful ways of using these solutions for computing various physical quantities of interest such as the jamming coverage, 
the mean number density, the mean cluster size, etc. 

\section{Numerical Simulations}

\begin{figure}[!htb]
\centerline{\includegraphics[width=8.0cm]%
{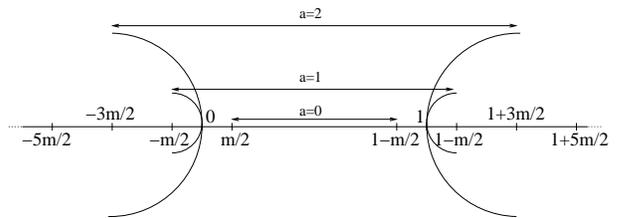}}
 \caption{Schematic illustration of the recursive simulation scheme.}
 \label{fig:3}
\end{figure}
To test the physical description of our CBD model, we have simulated it on a computer. One obvious constraint of the 
simulation is of course the finite size effect. However, for a sufficiently large substrate in comparison to the 
depositing particles, the finite-size effect can be made sufficiently small.  
To simplify the simulation, we use the approximation mentioned in the previous section that the gaps are 
uncorrelated and can be treated independently. This allows us to treat the problem in a recursive way if we focus on 
the coverage only and ignore the kinetics of the process. In brief,  the description of the recursive scheme of the 
simulation is as follows. Let us assume that the initial gap of unit interval $[0,1]$ bordered on both ends by semidisks 
of radius $(2a-1)m/2$ as shown in Fig. 3. We then generate a random 
number $n$ from the interval $[-(2a-1)m/2,1+(2a-1)m/2]$ to assign to the center of the incoming particle. The incoming 
particle is then adsorbed directly creating two new smaller intervals if $n$ lies in the 
interval $[m/2,1-m/2]$. Otherwise, the disk is adsorbed on one of the edges of $[0,1]$ creating only one new gap 
of interval $[m,1]$ or of $[0,1-m]$ depending on if is $n \in [-(2a-1)m/2,m/2]$ or $n\in [1-m/2,1+(2a-1)m/2]$ respectively.
% of size $[a+m/2,1]$ or $[0,1-(a+m/2)]$ depending
%on the random number if it is within $[0,a]$ or $[1-a,0]$ respectively. 
In the case when the disk is adsorbed in one of the two edges, we increase the 
counter which provides the information on the mean cluster size of the system. We then continue the process assuming
each new gap is again bordered by virtual semidisk and treat them in the same fashion as for the first step until we 
have no more gaps of size $\geq m$. At this point we add all the 
gaps of size $< m$ and finally, using this, we can immediately calculate the jamming coverage and the mean cluster size.
We have performed the simulation with substrate size $\sim 10^6m-10^8m$ within the interval $[0,1]$ and found an 
excellent match with the corresponding analytical results up to a several digits. We also noticed that increasing the 
substrate size by decreasing the $m$ value only contributes to a higher order precision as expected. 

%We have a gap of size $x$ which is defined as an interval having half of a disk of diameter $m$ at its two remote ends.
%However, for the CBD model before tossing for the random number we replace the disks that define the given gap $x$ by 
%the virtual disks of diameter $(2a-1)m$ keeping the gap size intact (see Fig. 1). 
%The first step can be described as follows. We find 
%a random position from the length $x+(2a-1)m$ to assign to the center of the incoming article. The incoming particle is 
%then adsorbed directly creating two new smaller gaps if the chosen position lies within  $(x-m)$ as shown in Fig. 1.
%Otherwise, the disk is adsorbed via rolling creating only one new gap $(x-m)$. In the latter case, we increase the 
%counter which can provides us the information on mean cluster size of the system. We continue the process treating each 
%new gap in the same fashion as we described for
%the first step until we have no more gaps of size $x\geq m$. At this point we add all the 
%gaps of size $\leq m$ and finally using this we can immediately calculate the jamming coverage.

\section{Results and analysis}

The fraction of the line covered by the adsorbing particles or 
the coverage $\theta(a,t)$ at different instants of time can be defined as
\begin{equation}
\theta(a,t)=1-\int_0^\infty xc(x,t)dx,
\end{equation}
while the number density is defined by the following relation
\begin{equation}
N(a,t)=\int_0^\infty c(x,t)dx.
\end{equation}
However, we find it more convenient to handle their rate equation rather than their definition itself. 
The kinetic equation for the coverage is  
\begin{equation}
\label{cov}
{{d\theta(a,t)}\over{dt}}=m\int_m^\infty\{x+m(2 a-1)\}c(x,t)dx =\Phi(a,t).
\end{equation}
Here, the quantity $\Phi(a,t)$ is the fraction of the substrate accessible to a new particle at a given time $t$.
The kinetic equation for the number density on the other hand is
\begin{equation}
\label{num}
{{dN(a,t)}\over{dt}}=\int_m^\infty (x-m)c(x,t)dx.
\end{equation}
The above two equations can be combined together to obtain
\begin{equation}
{{d\theta(a,t)}\over{dt}}=m{{dN(t)}\over{dt}}+2am\int_m^\infty c(x,t)dx.
\end{equation}
Note that in the simple RSA, one gap corresponds to one particle 
and thus we have $\theta(0,t)=m N(0,t)$ reflecting the fact that the average particle size is the same as the size of
the adsorbing particles. However, in the present case, for $a>0$, the second term 
of the above equation describes the cluster formation. That is, the first term on the right hand side of Eq. (13) 
takes into account the direct deposition while the effect of the rolling mechanism is described by the second term.
Using the solution for the appropriate boundary into Eq. (\ref{cov}) yields
\begin{equation}
\theta(a,t)=\int_0^{mt}F(a,u)(1+2au)du.
\end{equation}
This can provide all the information about kinetics aspect of the process, namely how the coverage evolves in time. 
%The fraction of the substrate accessible to a new arrival is $\Phi(a,0)=1$ and $\Phi(a,\infty)=1$ and at intermediate
%time $t$ it can be obtained from the follwoing relation,
%\begin{equation}
%\Phi(a,t)= F(a,mt)(1+2amt).
%\end{equation}
\begin{figure}[!htb]
\centerline{\includegraphics[width=80mm,height=50mm,angle=0]%
{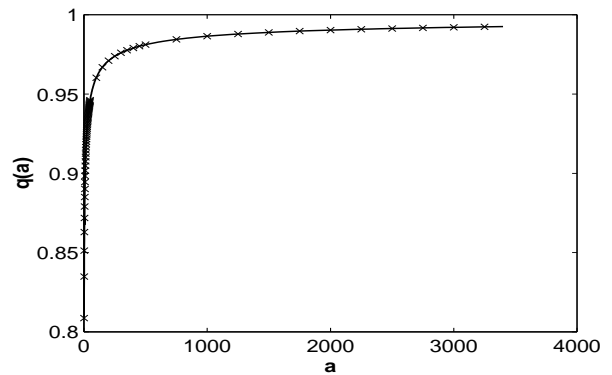}}
 \caption{Jamming coverage as a function of $a$ that measures the strength of attraction: $q(a)$ vs $a$.}
  \label{fig:4}
\end{figure}
\begin{figure}[!htb]
\centerline{\includegraphics[width=80mm,height=50mm,angle=0]%
{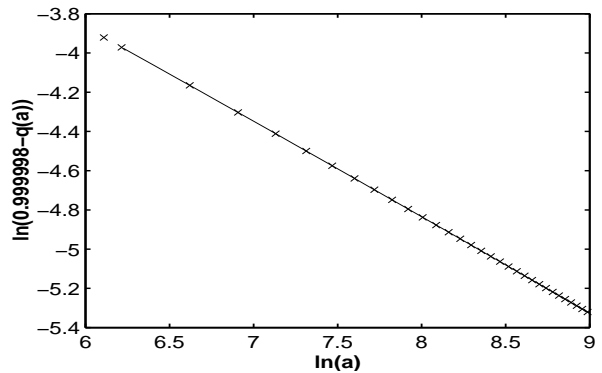}}
\caption{The linear fits of $\ln(q(\infty)-q(a))$ vs $\ln(a)$ having slope $-1/2$ in the large $a$ regime reveals 
that the convergence of the jamming coverage towards the closest packing is power-law in nature.}
  \label{fig:5}
\end{figure}
%\begin{figure}[!htb]
%\centerline{\includegraphics[width=80mm,height=60mm,angle=0]%
%{cov_vs_a.eps}}
% \caption{Dependence of coverage on the distance the precursor species can travel: $q$ vs $a$.}
%  \label{fig:2}
%\end{figure}
%\begin{figure}[!htb]
%\centerline{\includegraphics[width=8.0cm]%
%{logcov_vs_loga.eps}}
% \caption{The convergence of $q(a)$ towards its asymptotic value: 
%$Y=\ln[0.999998-q(a)]$ vs $X=\ln[a]$.}
%  \label{fig:3}
%\end{figure}
One of the characteristics of the deposition process is that the system reaches a state of dead-lock 
in a finite time when particles can no longer be adsorbed. This is typically known as the jamming limit 
and the exact critical time to reach such a state should depend on $a$.
However, till to-date there do not exist any theoretical means to pin down the exact critical time for reaching the 
jamming limit. Nevertheless, we can safely calculate the coverage in the jamming limit as 
\begin{equation}
q(a)=\lim_{t\longrightarrow \infty} \theta(a,t).
\end{equation}
The jamming coverage has been of special interest in the study of the deposition phenomena as it can 
uniquely characterize the structure of the resulting monolayer.  From the exact expression for
the coverage, Eq. (15), it is of particular interest to know the approach of the coverage $\theta(a,t)$ to the 
corresponding jamming limit $q(a)$. We find that beyond the transient behavior, the system reaches its
corresponding asymptotic coverage namely the jamming limit exponentially, with a decay constant $2a$, 
multiplied by an algebraic prefactor $t^{-1}$
\begin{equation}
q(a)-\theta(a,t) \sim t^{-1}e^{-2at}, 
\end{equation}
which was also reported in \cite{kn.viot}. 
Obviously, the classical RSA ($a=0$), we recover the power-law behavior which is also known as 
Feder's law \cite{kn.feder1}. Here for $a>0$, the exponential approach towards the jamming limit reflects the fact that 
the increasing number of particles that land on an already adsorbed particles are successfully accommodated via 
rolling. Another interesting point to check is how the jamming limit varies as we increase the 
strength of interaction $a$. In other words, we want to see how the 
jamming limit changes as we increase the degree of correlation between the particles in the adsorbed and adsorbing phases.
Fig. 4 shows  a sharp rise in the jamming coverage at low $a$ and a slow rise towards 
the closest packing in the large $a$ regime. In an attempt to quantify the slow regime we plot 
$\ln(q(\infty)-q(a))$ against $\ln(a)$ in Fig. 5 and find that the jamming coverage converges towards the closest packing
obeying a power law 
\begin{equation}
q(\infty)-q(a) \sim a^{-1/2},
\end{equation}
where $q(\infty)\approx 1$. It is important to note here that the system never reaches a complete closest packing ($q=1$) 
even for $a\longrightarrow \infty$. This is due to the fact that the substrate size too is of the same order as that 
of the size of the virtual disk, hence there is always a non-zero probability for a direct deposition at least
in the early stage. We attempted to check it in the computer choosing both the virtual diameter and the initial 
substrate size to be of the same order and large enough to minimize the finite size effect. This is exactly the case 
described by the analytical model as we let $a\longrightarrow \infty$. In doing so we never find 
a cluster covering the whole substrate . We checked it over and over again by increasing the substrate size and
the virtual disk size up to $\sim 10^10m$. Nevertheless, neither analytical solution nor the simulation could give us 
an exact estimate for $q(\infty)$.
%\begin{figure}[!htb]
%\centerline{\includegraphics[width=8.5cm]%
%{assymp_coverage.eps}}
% \caption{The convergence of $q(a)$ towards its asymptotic value: 
%$Y=\ln[0.99998-q(a)]$ vs $X=\ln[a]$.}
%  \label{fig:2}
%\end{figure}
\begin{figure}[!htb]
\centerline{\includegraphics[width=80mm,height=50mm,angle=0]%
{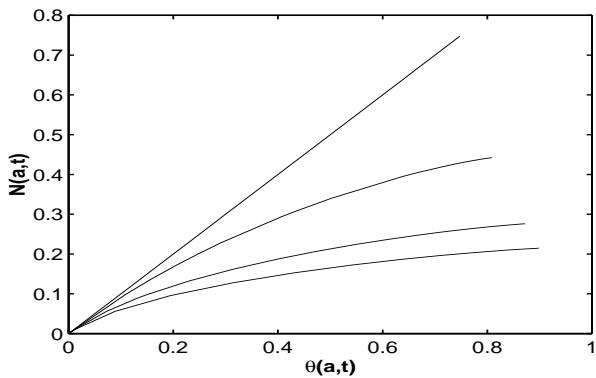}}
 \caption{Time dependence of the number density as a function of the coverage: The curves
from top to bottom represents $N(a,t)$ vs $\theta(a,t)$ for $a=0,1,5$ and $10$.}
  \label{fig:6}
\end{figure}

We now intend to obtain an exact expression for the number density by substituting the solution $c(x,t)$ for $x>m$
into Eq. (\ref{num}) which yields
\begin{equation}
N(a,t)={{\int_0^{m t} F(a,u)du}\over{m}}.
\end{equation}
%\begin{figure}[!htb]
%\centerline{\includegraphics[width=75mm,height=50mm,angle=0]%
%{a_k.eps}}
%\caption{Mean cluster size $k$ against $a$.}
%  \label{fig:6}
%\end{figure}
\begin{figure}[!htb]
\centerline{\includegraphics[width=80mm,height=50mm,angle=0]%
{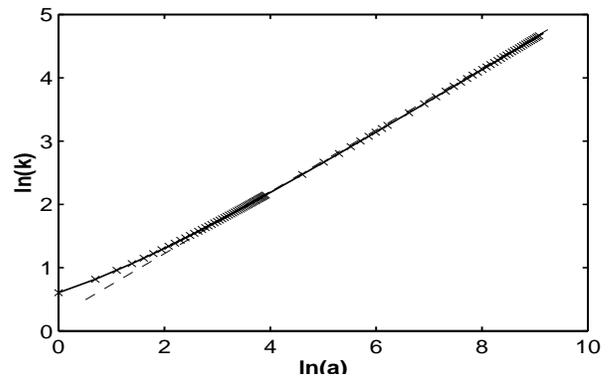}}
\caption{The linear fits of the plot of $\ln(k)$ vs $\ln(a)$ in the large $a$ regime with slope $1/2$ shows
that the mean cluster size increases as $\sim a^{1/2}$.}
  \label{fig:7}
\end{figure}
%\begin{figure}[!htb]
%\centerline{\includegraphics[width=8.0cm]%
%{logacs_vs_loga.eps}}
%\caption{Mean cluster size $s$ vs $a$ in the jamming limit for generalized ballistic deposition with $m=1$.}
%  \label{fig:6}
%\end{figure}
The above relation for the number density immediately implies that it depends on the size of the adsorbing particles. 
Here it is noteworthy to mention that the mean number density for the classical RSA ($a=0$) is simply 
the coverage divided by the size of the adsorbing particles. In this case, the mean number density increases 
linearly with the coverage as time proceeds having a slope $m$ during the process (see Fig. 6). 
However, as soon as a particle that fell on a previously adsorbed one can 
roll over the latter, the linear relation between $N(a,t)$ and $\theta(a,t)$ is immediately ceased and it is
replaced by a non-linear relation (see Fig. 6). Therefore, the mere knowledge of one of 
the two is not sufficient to obtain the other. This is due to the 
fact that the mean cluster size is different from the size of the adsorbing particle as the system keeps producing 
connected clusters of different sizes depending on the value of $a$. Fig. 3 shows that for $a>0$ the number density
grows linearly at a very low coverage. This is due to the fact that, at an initial stage the incoming particles hardly 
encounter any pre-occupied species and therefore there exists almost no cluster. However, as the substrate gets crowded,
 it is evident from Fig. 6 that the mean number density increases 
in a nonlinear fashion and the strength of non-linearity increases with increasing $a$. Therefore, to obtain the coverage 
(the number density) from the number density (coverage) we need to know the mean cluster size. 
%\begin{figure}[!htb]
%\centerline{\includegraphics[width=8.5cm]%
%{cluster_size_t.eps}}
% \caption{Mean cluster size $<l(t)>$ vs $t$ for generalized ballistic deposition with $m=1$.}
%  \label{fig:2}
%\end{figure}
The expression for the coverage $\theta(a,t)$ and  the mean number density $N(a,t)$  at different 
instants of time can give us an estimate of how the mean cluster size defined as
\begin{equation}
s(t)={{\theta(a,t)}\over{N(a,t)}}
\end{equation}
grows in time and with the strength of $a$. We find that for a given size of the adsorbing species, 
the mean cluster size in the jamming limit is 
\begin{equation}
\lim_{t \longrightarrow \infty}{{\theta(a,t)}\over{N(a,t)}}=s={{\int_0^{\infty}F(a,u)(1+2au)du}\over
{\int_0^{\infty} F(a,u)du}}m.
\end{equation}
%The above relation shows that for a given $m$, the mean cluster size increases monotonously and remains as a constant
%quantity.

Obviously, like the mean number density,  the mean cluster size should depend on the 
size of the adsorbing particle. However, the ratio between the two $k=s/m$ remains constant in the jamming limit and
therefore it is useful to call it the universal mean cluster size. It is note worthy to mention here that for the simple 
RSA we get $k=1$ and hence the mean cluster size is the same as that of the size of the 
adsorbing species. However, for $a>0$ we find that the universal mean cluster $k>1$ and $k$ increases  
monotonously with increasing $a$. Furthermore, Fig. 7 reveals that in the large $a$ regime the universal mean 
cluster size $k$ increases as
\begin{equation}
k\sim a^{1/2}.
\end{equation} 
The probability of adsorption of particles without overlapping with any preadsorbed particle decreases with
time and for all $a$; however the strength of decrease gets sharper and sharper as $a$ increases. 
For $a \longrightarrow \infty$, only at a very
initial stage some particles may be adsorbed by direct deposition without overlapping. Thus, the system 
never reaches a state of closest packing but of almost closest packing. This is well supported by our numerical
simulation.     
The jamming coverage thus increases with increasing $k$ (see Fig. 8). We find that like 
the $q(a)$ vs $a$, the approach of the jamming limit towards almost closest packing against $k$ also follows
a power law form but with a different exponent. As shown in Fig. 9, the plot of $q(k(a\longrightarrow \infty))-q(k)$
versus $k$ in the logarithmic scale along both axis is well fitted by a straight-line with slope $1$ and hence 
\begin{equation}
q(k(a\longrightarrow \infty))-q(k) \sim k^{-1},
\end{equation}
where $q(k(a\longrightarrow \infty)) \approx 1$.
\begin{figure}[!htb]
\centerline{\includegraphics[width=80mm,height=50mm,angle=0]%
{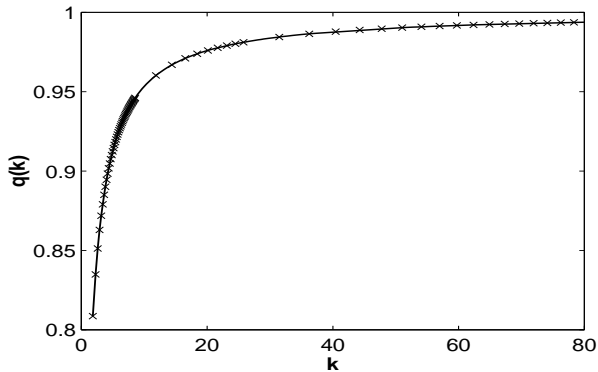}}
\caption{The jamming coverage as a function of the mean cluster size $k$.}
  \label{fig:8}
\end{figure}
\begin{figure}[!htb]
\centerline{\includegraphics[width=80mm,height=50mm,angle=0]%
{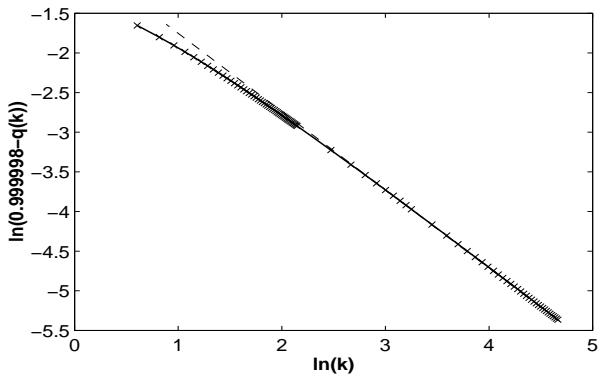}}
\caption{The linear fits of $\ln(q(\infty)-q(a))$ vs $\ln(k)$ in the large $k$ regime shows that the jamming 
coverage increases in a power-law fashion $\sim k^{-1}$.}
  \label{fig:9}
\end{figure}

%\begin{figure}[!htb]
%\centerline{\includegraphics[width=8.0cm]%
%{cov_vs_acs.eps}}
% \caption{Dependence of the jamming coverage on the mean cluster size in the same limit: $q(k)$ vs $k$.}
%  \label{fig:7}
%\end{figure}
%\begin{figure}[!htb]
%\centerline{\includegraphics[width=8.0cm]%
%{logcov_vs_logacs.eps}}
% \caption{Dependence of the jamming coverage on the mean cluster and its asymptotic behavior: 
%$Y=\ln[0.999998-q(k)]$ vs $X=\ln[k]$.}
%  \label{fig:8}
%\end{figure}

%\begin{figure}[!htb]
%\centerline{\includegraphics[width=8.5cm]%
%{mean_size.eps}}
% \caption{Mean cluster size $<l>$ vs $a$ in the jamming limit for generalized ballistic deposition with $m=1$.}
%  \label{fig:1}
%\end{figure}

\section{Discussion}

We first discuss various possible interpretations for different regimes constituted by different values of $a$
in our CBD model.
It is interesting to note that when $a=1$, the size of the virtual and the real disk coincides and hence every point along
the substrate has the equal probability of occupation by the incoming particle. In this case, 
particles that fall on an already adsorbed one may travel up to a linear distance of its own size $m$ 
via the rolling motion following the steepest descent path. The particle can either be trapped in the local
minimum or it can reach the global minimum. In the former case, it is rejected while in the latter case it
is adsorbed irreversibly touching the one it just rolled over, provided there is at least a gap to accommodate
it. The situation for $0<a<1$ is also interesting as it describes a mixed process composed of the simple RSA and the 
BD process. In this case, whenever an incoming disk encounters 
another disk, it is then allowed to roll over the latter disk with probability $a$.
The rolling mechanism over the previously adsorbed disk can be assumed to be due to the 
gravitational pull towards the adsorbing surface. The ballistic deposition model ($0<a\leq 1$) can 
thus describe the situation where the transport of the adsorbing species is dominated by gravitational effects. However,
the model with $1<a<\infty$ can describe the situation where the gravitational effects as well as the attractive 
interaction between the elements of the adsorbed phase and the incoming particles play the dominant role.

It is noteworthy to mention that Viot {\it et al} also generalised the  ballistic deposition model 
and gave the following definition  \cite{kn.viot}.
The disks of unit diameter are dropped uniformly and sequentially one at each time step. The disks can either reach 
the adsorbing plane or fall on an already adsorbed disk or on a cluster of disks. In the former case, the trial attempt 
is retained with a probability $q=(1-p)$. In the latter case, on the other hand, the trial disk follows the path of the 
steepest descent over the disk it encountered. The disk is then adsorbed with probability $p$ 
provided the particle can reach the global minimum by the roll over motion; otherwise it is trapped in an elevated 
position and it is rejected. In \cite{kn.viot} the parameter $a$ is defined as $a=p/q$. 
%For example, if we choose
%the case $a=2$ it would mean that the disk that could reach the surface without overlapping is adsorbed with probability
%$q=1/3$ and the one that reaches the surface via rolling is adsorbed with probability $p=2/3$. 
Therefore, the direct
adsorption and the deposition via the rolling mechanism  will both cause an increase in the rejection rate of the trial 
attempts for deposition. As a result, the approach to the jamming coverage in time should be slower as $a$ increases. 
However, the analytical solutions reveal an opposite behaviour, see Eq. (17). 

Note that it is the rejection criterion 
that determines how fast or how slow the coverage should evolve and finally reach the jamming limit. We would like to 
point out here that the apparent ambiguity arises because one cannot define  $a=p/q$ due to the following reasons.
We can rewrite Eqs. (1) and (2) for  $a=p/(1-p)$ upon multiplying both equations by $(1-p)$. 
%\begin{eqnarray}
%{{\partial c(x,t)}\over{\partial t}}& = &(1-p)\Big \{-(x-m)c(x,t)+2\int_{x+m}^\infty c(y,t)dy\Big \} \nonumber \\ && +
%2pm \Big \{c(x+m,t)-c(x,t)\Big\}
%\end{eqnarray}
%for $x\geq m$ and
%\begin{equation}
%{{\partial c(x,t)}\over{\partial t}} = 2 (1-p)\int_{x+m}^\infty c(y,t)dy +2pm c(x+m,t)
%\end{equation}
%for $x<m$.
If we then set $p=1$ in these equations we find that the remaining terms alone are incapable of describing any 
meaningful physical process. The point to emphasize here is that the roll over mechanism comes after the trial attempt for
deposition is made. The trial attempt to deposit the particle
is the primary event at each time step of the process which may then be followed by the roll over motion upon collision.
Therefore, the two events, the direct adsorption and the deposition via rolling are not mutually exclusive.

The model we have presented in this article is solved exactly 
by means of a kinetic equation approach and supported by the numerical simulations based on the recursive algorithm. 
The exact match between the analytical solution and the numerical simulation helped not only to  confirm the validity 
of the mean-field approximation  but also to shed a deeper insight into the nature of the problem.
The basic principle of the model is the same as that of the simple ballistic deposition process. We have 
extended the simple BD by adding a certain degree of correlation 
between the adsorbing particles and those already adsorbed. To increase the flexibility of the model, we allowed a  
parameter $a$ that can tune the strength of the correlation which is induced by the attractive force. 
Instead of using the attractive force directly, we have shown a way of transforming it into a virtual situation, 
which is then just the simple BD model. 
  
The most significant consequence of the presence of the attractive force is that it results in an increased packing 
fraction and the mean clsuter size due to the formation of the higher order connected clusters. 
Moreover, the jamming coverage increases with the increasing degree of strength of the attractive force.
Similar results has also been recently reported by Pastor-Satorras and Rubi \cite{kn.rubi}, who 
studied a model of correlated sequential adsorption by numerical simulation both in one and two dimensions. 
However, unlike the square well type PD studied here, they used a Gaussian and exponential type PD around the center 
of each adsorbed particle. Nevertheless, despite the apparent differences in the detailed nature of the 
forces or in the PD,
the qualitative behaviour seems remarkably identical to what we have found in this article. 
Pastor-Satorras and Rubi too observed the similar trend of the increase of jamming
coverage as well as the mean cluster size. In addition, they too reported the approach towards the closest packing 
in the limit where the correlation is maximal. This reveals that the qualitative effect of the attractive force is 
insensitive to the detailed nature of the force. However, in addition to solving the model analytically  
we were able to quantify the effect of the attractive force. To this end, we have shown that in the strong force 
regime the convergence towards the closest packing against the strength of the 
attractive force follows a power-law relation $q(\infty)-q(a)\sim t^{-1/2}$. Perhpas, the emergence of such a power-law 
behaviour implies a universal nature of the phenomena including the exponent in the sense that it is independent of 
the detailed nature of the attractive force. However, at this point it is just a conjecture and we intend to investigate 
it in our future work.

\section{Conclusion}

In this article, we have presented an extension of the simple BD model by incoporating an attractive force betweent the
elements of the adsorbed phase and the incoming particle. 
The most significant consequence of the presence of the attractive force is the increase in packing fraction or the
jamming coverage of the resulting monolayer as we increase the strength of the attractive force $a$. 
This is manifested through the increase in the mean cluster size $k$ and in the strong force regimes it increases
as  $\sim a^{1/2}$. We have shown that the system exhibits a power-law approach of the jamming coverage towards the 
closest packing both in $a$ and $k$ but with different exponents $\sim a^{-1/2}$ and $\sim k^{-1}$ respectively, 
except for the weak field. It is important to note that although the jamming coverage increases 
with the degree of correlation {\it vis-a-vis} the mean cluster size, we can never create one single connected cluster 
spanning the whole substrate and giving the coverage $q=1$ which is indeed typical to 1D problem. 
Nevertheless, it indicates the potential structural phase transition in heigher dimensions and it has been indeed 
observed in \cite{kn.rubi}.
 
Finally, as our model produces connected clusters of different sizes, it may well explain some aspects
of RSA of polydisperse mixture in some sense. However, the mixture of particles has the following restrictions. 
Namely, the mixture contains a strict lower and upper cut-off and all the particles of 
intermediate size are of the integral multiple of the smallest particle in the mixture. 

\acknowledgements

This work was supported by the Alexander von Humboldt-Foundation (M.\ K. \ H.) and
the Deutsche Volkswagen-Stiftung (N.\ W.).

%\begin{figure}[!htb]
%\centerline{\includegraphics[width=8.5cm]%
%{assymp_coverage.eps}}
% \caption{Asymptotic approach of mean cluster size. The top curve represents 
%$Y=\log[t(\theta(\infty)-\theta(t))]$ vs $t$
%and the bottom curve represents $Y=\log[\theta(\infty)-\theta(t)]$ vs $t$ with $m=1$.}
%  \label{fig:3}
%\end{figure}

\end{document}